\documentclass[12pt]{article}

\ifx\pdfoutput\undefined
\usepackage[dvips,bookmarks]{hyperref}
\else
\usepackage{hyperref}
\fi
\hypersetup{colorlinks=false,bookmarksopen,bookmarksnumbered,citecolor=blue,
   pdfstartview=FitH}

\usepackage[dvips]{graphicx}
\usepackage{latexsym}
\usepackage{amssymb,amsfonts,amsmath}
\usepackage{graphicx}
\usepackage{indentfirst}
 \usepackage{bbm}
\usepackage{mathrsfs}

\usepackage{amsmath, amsthm,amssymb}
\usepackage{mathrsfs}
\usepackage{hyperref}
\usepackage{amsfonts}
\usepackage{dsfont}
\usepackage{slashed, tensor}
\usepackage{booktabs}
\usepackage{graphicx}
\usepackage{mathrsfs}

\parskip=\medskipamount

\usepackage{lipsum}       
\usepackage{xargs}            

\usepackage[colorinlistoftodos,prependcaption,textsize=tiny]{todonotes}
\newcommandx{\unsure}[2][1=]{\todo[linecolor=red,backgroundcolor=red!25,
bordercolor=red,#1]{#2}}
\newcommandx{\change}[2][1=]{\todo[linecolor=blue,
backgroundcolor=blue!25,bordercolor=blue,#1]{#2}}
\newcommandx{\STinfo}[1]{\todo[backgroundcolor=red!25,bordercolor=red,noline]{S.T.:#1}}
\newcommandx{\SKinfo}[1]{\todo[backgroundcolor=blue!25,bordercolor=blue,noline]{S.K.:#1}}

\newcommand {\cD}{{\cal D}}

\newcommand {\cK}{{\cal K}}

\newcommand {\cM}{{\cal M}}
\newcommand {\cN}{{\cal N}}

\newcommand {\cR}{{\cal R}}

\newcommand {\cT}{{\cal T}}

\def\a{\alpha}

\def\b{\beta}
\def\c{\chi}
\def\d{\delta}

\def\g{\gamma}
\def\G{\Gamma}

\def\l{\lambda}
\def\m{\mu}

\def\q{\theta}

\def\s{\sigma}

\def\x{\xi}
\def\z{\zeta}

\def\F{\Phi}

\def\L{\Lambda}
\def\O{\Omega}

\def\U{\Upsilon}

\def\rd{{\rm d}}
\def\ri{{\rm i}}

\newcommand{\ad}{{\dot{\alpha}}}                           
\newcommand{\bd}{{\dot{\beta}}}                            
\newcommand{\ve}{\varepsilon}                            

\newcommand{\ab}{{\a\b}}

\newcommand{\pa}{\partial}                           
\newcommand{\hf}{\frac12}

\newcommand{\be}{\begin{equation}}
\newcommand{\ee}{\end{equation}}
\newcommand{\bea}{\begin{eqnarray}}
\newcommand{\eea}{\end{eqnarray}}

\newcommand{\ba}{\begin{array}}
\newcommand{\ea}{\end{array}}

\newcommand{\bm}[1]{\mbox{\boldmath$#1$}}

\def\double #1{#1{\hbox{\kern-2pt $#1$}}}

\newcommand{\hal}{{\hat{\a}}}

\newcommand{\gd}{{\dot\g}}

\newcommand{\sSU}{\mathsf{SU}}

\newcommand{\sSO}{\mathsf{SO}}

\newcommand{\bsubeq}{\begin{subequations}}
\newcommand{\esubeq}{\end{subequations}}

\numberwithin{equation}{section}

\begin{document}

\begin{titlepage}
\begin{flushright}
April, 2016 \\
\end{flushright}
\vspace{5mm}

\begin{center}
{\Large \bf 
The conformal supercurrents in diverse dimensions 
and conserved superconformal currents
}
\\ 
\end{center}

\begin{center}

{\bf
Yegor Korovin${}^{a}$,   
Sergei M. Kuzenko${}^{b}$ and  Stefan Theisen${}^{a}$ 
} \\
\vspace{5mm}

\footnotesize{
${}^{a}${\it Max-Planck-Institut f\"ur Gravitationsphysik, Albert-Einstein-Institut,\\
Am M\"uhlenberg 1, D-14476 Golm, Germany}
}
\vspace{2mm}
~\\
\footnotesize{
${}^{b}${\it School of Physics M013, The University of Western Australia\\
35 Stirling Highway, Crawley, W.A. 6009, Australia}}  
~\\
\vspace{2mm}

\end{center}

\begin{abstract}
\baselineskip=14pt
Given a conserved and traceless energy-momentum tensor and a conformal 
Killing vector, one obtains a conserved current. We generalise this
construction to superconformal theories in three, four, five and six dimensions
with various amounts of supersymmetry by working in the appropriate superspaces.   
\end{abstract}

\vfill

\vfill
\end{titlepage}

\section{Introduction}

We consider conformal field theories in $d$-dimensional Minkowski space
${\mathbb M}^d$. A central feature of conformal field theories is the 
existence of a symmetric, traceless and conserved  energy-momentum tensor 
$T^{ab}$
\bea
T^{ab}= T^{ba}~, \qquad \eta_{ab}T^{ab}=0~, \qquad \pa_b T^{ab}=0~,
\label{1.1}
\eea
with $\eta_{ab} $ the Minkowski metric (of mostly plus signature). 
Let $\x= \x^a \pa_a$ be a conformal Killing vector field of ${\mathbb M}^d$, 
\bea
\pa_a \x_b + \pa_b \x_a = \frac{2}{d}
\eta_{ab} \pa_c \x^c
~.
\label{1.2}
\eea
As is well-known, for every conformal Killing vector $\x$ one can construct a conserved
current $V^a$ as
\bea
\label{cc}
V^a= T^{ab}\x_b\,.
\label{1.300}
\eea
Conservation of $V^a$, 
\bea
  \pa_a V^a =0~
  \label{1.3}
\eea
is a consequence of the conservation and tracelessness of $T^{ab}$ 
(which hold on-shell).  

In this note, we generalise the above construction, $(T^{ab}, \x^a) \to V^a$, 
to the case of superconformal field theories in diverse dimensions
in a manifestly supersymmetric way. 
In the supersymmetric generalisation 
the traceless energy momentum tensor $T^{ab}$ will be
embedded in the conformal supercurrent multiplet while the conserved conformal
current $V^a$ will turn into the conserved superconformal current 
embedded in a supermultiplet.

A supersymmetric analogue of the energy-momentum tensor is 
the supercurrent multiplet introduced by Ferrara and Zumino \cite{FZ} 
in the framework of four-dimensional (4D) $\cN=1$ Poincar\'e supersymmetry. 
A supersymmetric extension of the notion of conserved current is 
the conserved current multiplet introduced by Ferrara, Wess and Zumino \cite{FWZ}
in the 4D $\cN=1$ super-Poincar\'e case. 
The concepts of supercurrent and conserved current multiplet also
exist  for different types of supersymmetry and in spacetimes of  
dimension $d\neq 4$. 
By definition, the supercurrent is a supermultiplet containing
the energy-momentum tensor and the supersymmetry current(s), along with
some additional components such as the $R$-symmetry current.
In this note, we define the conserved current multiplet to be a supermultiplet 
containing a {\it single} conserved vector current (equivalently, a closed $(d-1)$-form), 
along with some other scalar and spinor components. 

Unlike the energy-momentum tensor, the functional structure of the supercurrent 
and the corresponding conservation equation depend on the dimension 
and supersymmetry type. As an example we recall the $\cN=1$ and 
$\cN=2$ supersymmetric extensions of the conformal energy-momentum tensor 
\eqref{1.1} in four dimensions.
The $\cN=1$ conformal supercurrent \cite{FZ} is a real vector
superfield $J_{\a\ad}$ constrained by 
\bea
D^\a J_{\a\ad} =0 \quad  \Longleftrightarrow \quad \bar D^\ad J_{\a\ad} =0~.
\label{1.5}
\eea
The $\cN=2$ conformal supercurrent \cite{SohniusN=2} is a real scalar 
superfield $J$ constrained by
\bea
D^{ij} J  = 0 \quad  \Longleftrightarrow \quad
{\bar D}^{ij} J = 0~,
\label{sccl}
\eea 
where $D^{ij} = D^{\a  i} D^{j}_\a =D^{ji}$, 
${\bar D}^{ij}
= {\bar D}^{i}_\ad {\bar D}^{j \ad } =\bar D^{ji}$. 
It is also pertinent to recall  the $\cN=1$ and 
$\cN=2$ supersymmetric extensions of the conservation equation
\eqref{1.3} in four dimensions. The $\cN=1$ conserved current multiplet 
is described by a real linear superfield $L$ \cite{FWZ}  constrained by 
\bea
D^2 L =0 \quad  \Longleftrightarrow \quad
\bar D^2 L=0~.
\label{1.77}
\eea
In the $\cN=2$ case, the conserved current multiplet is described by a real linear 
superfield $L^{ij}$ \cite{BS}, which is defined to be 
a real $\sSU(2)$ triplet $L^{ij} = L^{ji}$, 
$\overline{L^{ij}}=L_{ij}= \ve_{ik} \ve_{jl} L^{kl}$,
constrained by 
\bea
D^{(i}_\a {L}^{jk)} = 0 \quad  \Longleftrightarrow \quad
{\bar D}^{(i}_\ad {L}^{jk)}= 0~.
\eea

In order to introduce a supersymmetric analogue of the construction, 
$(T^{ab}, \x^a) \to V^a$, we also need a supersymmetric counterpart of  
the notion of conformal Killing vector. 
It was originally given by Sohnius \cite{Sohnius} in the case of 4D $\cN$-extended 
Poincar\'e supersymmetry
and further developed in four \cite{Sohnius,Lang,BPT,Shizuya,BK,HH,Park4D,KT}
and other \cite{GM,Park6D,Park3D,K-compactified,KPT-MvU} dimensions.

In the next section we review the definition of conformal Killing supervector fields. 
This is further elaborated in Appendix A. 
Sections 3 to 5 are devoted to superconformal theories in three, four, 
five and six  dimensions, 
respectively. In each of these sections we first review known results 
about the superconformal currents. We also state the general properties of 
conserved current multiplets and their conservation laws. 
We then construct these conserved currents in terms of the supercurrents and the 
conformal Killing supervector fields, generalising eq. \eqref{cc}.
Section 6 discusses modifications that occur for non-conformal supercurrents in 
curved backgrounds. 


\section{Conformal Killing supervector fields}

According to Nahm's classification \cite{Nahm}, 
superconformal algebras exist in spacetime dimensions $d \leq 6$.
In superspace, the superconformal transformations are generated by the 
so-called conformal Killing supervector fields. The latter can be defined in several 
different but equivalent ways. We first recall the least orthodox but, probably,  
most useful definition. We then present the more 
standard definition.

Let ${\mathbb M}^{d|\d}$ be a Minkowski superspace with $d\leq 6$ spacetime dimensions 
and $\d$ fermionic dimensions.
We denote by  $z^A = (x^a, \q^{\hat \a}) $ the bosonic ($x^a$)
and fermionic ($\q^{\hat \a}$) coordinates
for ${\mathbb M}^{d|\d}$. The index $\hat \a$ of the Grassmann 
variable is, in general,  composite; 
it is comprised of a spinor index $\bm \a$ and an $R$-symmetry index $I$.  
The superspace  covariant derivatives are 
$D_A = (\pa_a, D_{\hat \a}) = e_A{}^M \pa_M$
such that $\{ D_{\hat \a} , D_{\hat \b} \} = T_{\hat \a \hat \b}{}^c \pa_c$
and $[D_{\hat \a} , \pa_b ]=0$,  
where $T_{\hat \a \hat \b}{}^c$ is the flat-superspace torsion tensor, which is constant. 

A real even supervector field\footnote{A supervector field $\x$ is real 
and even if $\x \F$ is real and even 
for every real bosonic superfield $\F$. In what follows, all supervector fields will be 
real and even.}  
over  ${\mathbb M}^{d|\d}$,
$\x = \bar \x = \x^A  D_A = \x^a \pa_a +\x^{\hat \a} D_{\hat \a}$, 
is said to be conformal Killing if the following condition holds
\bea
[\x , D_{\hat \a} ] \propto D_{\hat \b}\quad 
\Longleftrightarrow \quad 
[\x , D_{\hat \a} ]  = - (D_{\hat \a} \x^{\hat \b} ) D_{\hat \b}~.
\label{def}
\eea
This condition implies that the only independent components of $\xi$
are the vector ones, $\xi^a$.  
The set of all conformal Killing supervector fields forms a superalgebra  
(with respect to the standard Lie bracket)
which is isomorphic to the superconformal algebra. 

The above definition was used to introduce the conformal Killing supervector fields 
in the 5D  \cite{K-compactified} and 3D $\cN$-extended \cite{KPT-MvU} 
cases. However, it  differs somewhat from that used in \cite{KT} in the 4D 
$\cN$-extended case, as a generalisation of the 4D $\cN=1$ analysis in \cite{BK}.
Let us show that the definition used in \cite{KT} follows from the 
one above.

The coordinates of 4D $\cN$-extended Minkowski superspace ${\mathbb M}^{4|4\cN}$ 
are $z^A = (x^a , \q^\a_i, \bar \q_\ad^i)$, where $\a$ and $\ad$ are two-component 
spinor indices, and $i=1, \dots , \cN$ is an $R$-symmetry index.\footnote{Our 4D 
notation and conventions follow \cite{BK,WB}.} 
The spinor covariant derivatives obey the anti-commutation relations
\bea
\{D_\a^i , D_\b^j \} =0~, \qquad \{\bar D_{\ad i} , \bar D_{\bd j} \} =0~,\qquad
\{D_\a^i , \bar D_{\bd j} \} = - 2\ri \, \d^i_j \pa_{\a \bd} ~,
\eea
where $\pa_{\a\bd} = (\s^c)_{\a \bd} \pa_c$.
Given a supervector field $\x =\x^a \pa_a +\x^\a_i D_\a^i 
+\bar \x_\ad^i \bar D_i^\ad$, 
the condition \eqref{def} implies
\bea
\bar D_{\ad i} \x^{\bd \b} = 4\ri \, \d_\ad^\bd \x^\b_i~.
\label{2.33}
\eea
Making use of this result, a short calculation gives
\bea
\bar D_{\ad i} \bar D_{\bd j} \x^{\gd \g } =0\quad \Longleftrightarrow \quad
\bar D_{\ad i} \x^\b_j =0~.
\eea
Equation \eqref{def} is thus equivalent to 
\bea
[\x, D_\a^i ] \propto D_\b^j~,
\eea
which is the definition of the conformal Killing supervector fields used 
in \cite{KT}.

Eq. \eqref{2.33} is equivalent to the relations 
\begin{subequations} \label{2.6ab}
\bea
\x^\a_i = -\frac{\ri }{8} \bar D_{\ad i} \x^{\ad \a } 
\quad & \Longleftrightarrow &\quad 
\bar \x_\ad^i =  -\frac{\ri }{8} D^{\a i} \x_{\a \ad } ~,
\label{2.6a}
\\
D^i_{(\a } \x_{\b) \bd} =0 \quad & \Longleftrightarrow &\quad 
\bar D_i^{(\ad} \x^{\bd) \b} =0 ~.
\label{2.6b}
\eea
\end{subequations}
Relations \eqref{2.6a} express the spinor components of $\x^A$ in terms of the vector 
one. Eq. \eqref{2.6b} is the 4D $\cN$-extended 
{\it superconformal Killing equation}, which is a supersymmetric counterpart of 
\eqref{1.2}. In conjunction with the definition \eqref{2.6a}, it proves to contain all the information about the conformal Killing supervector fields. An obvious consequence of
\eqref{2.6b} is that the 
vector superfield parameter  
$\x_{\b\bd} = (\s_a)_{\b\bd} \, \x^a(x,\q,\bar \q)$ obeys the equation
\bea
\pa_{(\a (\ad}\x_{\b) \bd)} =0~.
\label{2.77}
\eea
Switching off the Grassmann variables gives the vector field
$\x^a| := \x^{a} (x,\q , \bar \q) |_{\q= \bar \q =0}$,  which is 
an ordinary conformal Killing vector field. 
Indeed, \eqref{2.77} coincides with the $d=4$ conformal Killing equation  \eqref{1.2} rewritten 
in the two-component spinor notation.

The traditional definition of superconformal transformations in superspace 
was originally given by Sohnius \cite{Sohnius} in the 4D $\cN$-extended case.\footnote{Sohnius
 simply generalised the standard definition 
 of  infinitesimal conformal transformations of ${\mathbb M}^d$
as those transformations which at most scale the interval 
$\rd s^2 =\eta_{ab} \rd x^a \rd x^b$, where  $\eta_{ab}$ is the Minkowski metric.} 
Park used this definition to introduce
the superconformal transformations in the 6D $\cN=(p, 0)$ and $\cN=(0,q)$  \cite{Park6D} and 
3D $\cN$-extended \cite{Park3D}  Minkowski superspaces.
According to this definition, an infinitesimal coordinate transformation 
$\delta z^A = \x^A(z)$ generated by a  
supervector field $\xi^A$ on ${\mathbb M}^{d|\d}$,  
is called superconformal if it at most scales the supersymmetric interval 
$\rd s^2 = \eta_{a b} e^a e^b$. 
Here the supersymmetric one-forms \cite{AV}
$e^A = \rd z^M e_M{}^A$ constitute the dual basis for $D_A$ 
defined by $\rd =  \rd z^A \pa_A = e^A D_A$.
Using this definition, Park showed  \cite{Park6D} that  
in six dimensions superconformal transformations and hence superconformal algebras
exist only for the supersymmetry types 
$\cN=(p,0) $ and $\cN=(0,q)$. While one can define $\cN=(p,q)$ Poincar\'e supersymmetry 
for any non-negative integers $p$ and $q$, 
in the mixed case with $p,q\neq 0$, 
the most general conformal Killing supervector field describes only super-Poincar\'e, 
$R$-symmetry and scale transformations. Analogous considerations 
may be used to show that in five dimensions, where one can introduce $\cN$-extended 
Poincar\'e supersymmetry (with $8\,\cN$ supercharges),  
superconformal algebras exists only for $\cN=1$; 
see \cite{HL} for a recent derivation. 

It is an instructive exercise to show that invariance of 
$\rd s^2$ leads to \eqref{def} which in turn allows to express $\xi^{\hat a}$ 
in terms of $\xi^a$ which itself satisfies the conformal Killing equation.   
Equivalence of the two definitions of conformal Killing supervector fields   
may also be established using a more general (third) definition, 
which is reviewed in Appendix \ref{AppendixA}.  


\section{Superconformal theories in three dimensions} 

In this section we consider superconformal field theories in three dimensions. 
The 3D $\cN$-extended conformal supercurrents have been described in \cite{KNT-M}
using the conformal superspace formulation 
for $\cN$-extended conformal supergravity, which was
developed in \cite{BKNT-M}.  In the locally supersymmetric case, 
the supercurrent of a superconformal field theory coupled to 
conformal supergravity is characterised by the same superfield type and 
the differential constraints as the super-Cotton tensor, which is the only 
independent curvature tensor of $\cN$-extended conformal superspace \cite{BKNT-M}. 
Upon freezing the conformal superspace to 
$\cN$-extended Minkowski superspace ${\mathbb M}^{3|2\cN}$,
we end up with the conformal supercurrents which were discussed in detail 
in \cite{BKS1,BKS2} and which we are going to review.
Here we parametrise ${\mathbb M}^{3|2\cN}$ by real Cartesian coordinates
$z^A= (x^a, \q^\a_I)$, where $I=1,\dots , \cN$. 
The spinor covariant derivatives $D_\a^I$ 
on  ${\mathbb M}^{3|2\cN}$ satisfy the anti-commutation relation 
$\{D^I_\a,D^J_\b\}=2 \ri \delta^{IJ}\gamma^m_{\a\b}\partial_m$; 
see \cite{KPT-MvU} for further details. 

For every  $\cN=1,2\dots$, the conformal supercurrent is a {\it primary} real tensor 
superfield in the sense of the superconformal transformation law  (5.1) in \cite{BKS1}. 
The conformal supercurrents\footnote{The $\cN=2$ (non-)conformal supercurrents 
were studied in \cite{DS,KT-M11}.}
for $\cN\leq 4$ are specified by the following properties:
\bea
&&  \left|
\begin{array}{c || c |c|c}
\hline
\mbox{SUSY Type}  ~& ~\mbox{Supercurrent} ~&~\mbox{Dimension}~ &
~\mbox{Conservation Equation}  \\
\hline
\cN=1  ~&  J_{\a\b\g}~& 5/2& ~ D^\a J_{\a\b\g} =0
\\
\hline
\cN=2 ~&  J_{\a\b} ~ &2& ~ D^{I \a} J_{\a\b} =0 \\
\hline
\cN=3 ~&  J_{\a} ~ &3/2& ~ D^{I \a} J_{\a} =0 \\
\hline
\cN=4 ~&  J ~ &1&~ (D^{I \a }D_{\a}^J - \frac{1}{4}\d^{IJ}D^{L \a}D _{\a}^L )J =0\\
\hline
\end{array}
\right| ~~~~~
\label{3.1supercurrent}
\eea
For $\cN>4$, the conformal supercurrent is a completely antisymmetric
dimension-1 superfield, $J^{IJKL} = J^{[IJKL]} $, subject to the conservation equation
\bea
D_{\a}^I J^{JKLP} = D_\a^{[I} J^{JKLP]}
- \frac{4}{\cN - 3} D_{\a}^Q J^{Q [JKL} \d^{P] I} ~ .
\eea
In the $\cN=4$ case, this equation is identically satisfied for
$J^{IJKL} = \ve^{IJKL} J$. That is why the $\cN=4$ supercurrent $J$ obeys
the second-order conservation equation given in \eqref{3.1supercurrent}. 

In three dimensions, one may think of a conserved current $V^a$, eq. \eqref{1.3}, 
as the Hodge dual of the gauge-invariant field strength $F = \rd A$ 
of a gauge one-form $A$. 
For this reason an $\cN$-extended conserved current multiplet 
may be characterised by the same superfield type and 
the differential constraints as the field strength of an $\cN$-extended 
Abelian vector multiplet \cite{Siegel,HitchinKLR,ZP,ZH,KLT-M11}. 
The conserved current multiplets with $\cN \leq 4 $ were reviewed 
in \cite{BKS1,BKS2}. 
In the $\cN=1$ case, the conserved current multiplet is a real spinor superfield
$L^\a$  constrained by 
\bea
D^\a L_{\a} =0~.
\label{3.3}
\eea
For $\cN>1$, it is a real antisymmetric superfield,
$L^{I J} = - L^{JI}$, constrained by 
\bea
D_{\a}^{I} L^{ J K}&=&
D_{\a}^{[I} L^{ J K]}
- \frac{2}{\cN-1}  D_{\a }^L L^{ L[J} \d^{K] I}
\label{3.4}
~.
\eea
This equation is identically satisfied in the $\cN=2$ case for which 
$L^{IJ}= \ve^{IJ} L$. For $\cN=2$ the conserved current multiplet 
obeys instead the conservation equation
\bea
(D^{\a I}D_{\a}^J - \frac{1}{2}\d^{IJ}D^{K \a}D _{\a}^K )L =0~.
\eea
When $\cN=3$, it is more convenient to work with the Hodge dual of $L^{IJ}$, 
which is $L^I=\hf \ve^{IJK}L^{JK}$. The latter obeys the constraint 
\bea
D_\a^{ (I}L^{J)} - \frac{1}{3}\d^{IJ}D^{K }_\a L^K  =0 ~,
\eea
which is equivalent to \eqref{3.4} with $\cN=3$. 

The $\cN=4$ case is very special. Given an $\cN=4$ conserved 
current multiplet $L^{IJ}$, it can be uniquely represented as 
$ L^{IJ} = L^{IJ}_+ + L^{IJ}_-$, where $L^{IJ}_+$ and $L^{IJ}_-$ are self-dual
and anti-self-dual, respectively,
\bea
 \hf \ve^{IJKL}L^{KL}_\pm  = \pm L^{IJ}_\pm~.
 \eea
It turns out that each of  $L^{IJ}_+$ and $L^{IJ}_-$ obeys the conservation equation 
\eqref{3.4} with $\cN=4$. Thus there are two inequivalent current multiplets
in the $\cN=4 $ case. This is in accord with the fact that the $\cN=4$ $R$-symmetry 
group factorises, due to the the isomorphism
$\sSO(4) \cong  \big( {\sSU}(2)_{\rm L}\times {\sSU}(2)_{\rm R} \big)/{\mathbb Z}_2$.

For $\cN>4$, it turns out that the off-shell multiplet constrained by \eqref{3.4} 
possesses more than one conserved current at the component level.  
Moreover, it also  contains higher spin conserved 
currents for $\cN>5$ \cite{GGHN,KN}. Therefore, there
is no conserved current multiplet for $\cN>4$ in the sense of the definition 
given in Section 1.  

In the cases $\cN=2,3,4$, it is often convenient to switch from the real basis 
for the Grassmann variables $\q^\a_I$ to a complex one in accordance with the rules
described in \cite{KPT-MvU}. Schematically, this amounts to replacements: 
(i) $D_\a^I \to ( D_\a \, , \bar D_\a) $ for $\cN=2$; 
(ii) $D_\a^I \to D_\a^{ij} = D_\a^{ji}$   for $\cN=3$, where $i,j =1,2$; 
(ii) $D_\a^I \to D_\a^{i \bar i}$ for $\cN=4$, where $i, \bar i =1,2$.
We will use such types of parametrisation  in the remainder of this section, where we
discuss the conserved currents for $\cN=1,2,3,4$ in turn.

\subsection{$\cN=1$ superconformal symmetry}

Any supervector field $\xi$ on $\cN=1$ 
Minkowski superspace ${\mathbb M}^{3|2}$
has the expansion
\bea
\x = \x^A D_A = \x^a \pa_a + \x^\a D_\a~,
\eea
with the vector $\x^a$ and spinor $\x^\a$ components being real. 
Requiring $\x$ to be conformal Killing, eq.  \eqref{def},  
leads to the following conditions:
\begin{subequations}\label{2.2}
\bea
& &\x^\a = \frac{\ri}{6} D_\b \x^{\b\a}~,  \\
&& D_{(\g} \x_{\a\b)} =0~,
\label{2.2b}
\eea 
\end{subequations}
of which \eqref{2.2b} is the $\cN=1$ superconformal Killing equation.
With the help of the conformal supercurrent $J^{\a\b\g}$, 
which satisfies the conservation equation
\bea
D_\g J^{\a\b\g}=0 \quad \Longrightarrow \quad
\pa_{\b\g} J^{\a\b\g} =0~.
\label{2.1}
\eea
we construct the following spinor superfield: 
\bea
L^\a = -\hf \x_{\b\g} J^{\a\b\g}~.
\label{3.111}
\eea 
It follows from \eqref{2.2b} and  \eqref{2.1} that $L^\a$ obeys equation 
\eqref{3.3}, 
which defines a conserved current multiplet.

A few words are in order about the component structure of $J^{\a\b\g}$ and $L^\a$.
As follows from  \eqref{2.2b}, the supercurrent has two independent real component
fields, which are:
\bea
S^{\a\b\g} := J^{\a\b\g}| ~, \qquad T^{\a\b\g\d} :=  D^\d J^{\a\b\g}| = T^{(\a\b\g\d)}~,
\eea
where the bar-projection means, as usual, that the Grassmann variables 
must be switched off. 
Here $S^{\a\b\g} $ is the supersymmetry current, and $T^{\a\b\g \d} $
the energy-momentum tensor. Both currents are conserved, 
\bea
\pa_{\a\b} S^{\a\b\g} =0~, \qquad \pa_{\a\b} T^{\a\b\g\d} =0~.
\eea
Switching to the three-vector notation,  
$T^{\a\b\g \d} \to T^{ab} = \frac{1}{4} (\g^a)_{\a\b} (g^b)_{\g\d} T^{\a\b\g \d}= T^{ba}$
 and $S^{\a\b\g}\to S^{a \g}= -\hf (\g^a)_{\a\b} S^{\a\b\g}$, 
the energy-momentum is automatically traceless, $\eta_{ab}T^{ab}=0$,
and so is the supersymmetry current, $\g_a S^{a } =0$.

Given a conserved current multiplet $L^\a$ constrained by \eqref{3.3}, 
it has two independent real component fields, which can be chosen as 
\bea
\l^\a := L^\a|~, \qquad 
V^{\a\b} := D^\b L^\a | = V^{\b\a}~.
\eea
The vector field is conserved,
\bea
\pa_{\a\b} V^{\a\b} =0~. 
\eea
To compute the conserved current contained in \eqref{3.111}, 
one needs the explicit expression for an arbitrary  $\cN=1$ conformal Killing supervector 
field. The most general $\cN$-extended conformal Killing supervector field is given 
by eq. (4.4) in \cite{BKS1}. 


\subsection{$\cN=2$ superconformal symmetry}

Any supervector field $\xi$ on $\cN=2$ 
Minkowski superspace ${\mathbb M}^{3|4}$
has the form 
\bea
\x= \x^A D_A = \x^a \pa_a + \x^\a D_\a +\bar\x_\a \bar D^\a~,
\eea
where $\x^a$ is real, and $\bar \x^\a $ is the complex conjugate of $\x^\a$. 
Requiring $\x$ to be conformal Killing, eq.  \eqref{def},  
gives
\begin{subequations}\label{3DN=2CKSV}
\bea
 \x^\a &=& - \frac{\ri}{6} \bar D_\b \x^{\b\a}~, \\
D_{(\g} \x_{\a\b)} &=& \bar D_{(\g} \x_{\a\b)} =0 
\quad \Longrightarrow \quad D^2 \x_{\ab} = \bar D^2 \x_{\a\b} =0~.
\label{3DN=2CKSV.b}
\eea
\end{subequations}
Here \eqref{3DN=2CKSV.b} is the $\cN=2$ superconformal Killing equation. 
Together with the $\cN=2$ conformal supercurrent $J^{\a\b}$, which satisfies 
\bea
D_\b J^{\a\b}=
\bar D_\b J^{\a\b}=0 \quad \Longrightarrow \quad \pa_{\a\b} J^{\a\b}=0~,
\label{3DN2supercurrent}
\eea
we construct the scalar superfield
\bea
L= -\hf \x_{\a\b} J^{\a\b} = \x_a J^a~.
\label{3.199}
\eea	
It follows from \eqref{3DN=2CKSV.b} and
\eqref{3DN2supercurrent} 
that $L$ is a linear superfield, 
\bea
D^2 L = \bar D^2 L =0~,
\label{3.200}
\eea 
and therefore $L$ contains a conserved current. 

Because of the constraints \eqref{3DN2supercurrent}, 
the supercurrent has four independent component fields, which are
\bea
j^{\a\b}:= J^{\a\b}|~, \qquad S^{\a\b\g}:=D^\g J^{\a\b} | = S^{(\a\b\g)}~, 
\qquad T^{\a\b\g\d} := [ D^{(\g} , \bar D^{\d}] J^{\a\b)}|~,
\eea
as well as $\bar S^{\a\b\g}$, the complex conjugate of $S^{\a\b\g}$.
Here $j^{\a\b}$ is the $R$-symmetry current, $S^{\a\b\g}$ and $\bar S^{\a\b\g}$
the supersymmetry currents, and $T^{\a\b\g \d} $ the energy-momentum tensor. 
All these currents are conserved, as a consequence  of the constraints \eqref{3DN2supercurrent}.

Given a conserved current multiplet $L = \bar L$ constrained by \eqref{3.200}, 
it has five independent components, which can be identified with
\bea
l:= L|~, \qquad 
\l^\a := D^\a L| ~\qquad U:= \ri D^\a \bar D_\a L|~, \qquad 
V^{\a\b} := [D^{(\a}, \bar D^{\b)} ] L |~, 
\eea
as well as $\bar \l^\a$, the complex conjugate of $\l^\a$. 
The vector field is conserved, $\pa_{\a\b} V^{\a\b}=0$, as a consequence
of the identity 
\bea
[D^2 , \bar D^2] = -4 \ri \pa_{\a\b} [D^\a, \bar D^\b] ~.
\eea
To compute the conserved current contained in \eqref{3.199}, 
one has to make use of the explicit expression for the 
most general $\cN=$ conformal Killing supervector field given in \cite{BKS1}.


\subsection{$\cN=3$ superconformal symmetry}

Any supervector field $\xi$ on $\cN=3$ 
Minkowski superspace ${\mathbb M}^{3|6}$
has the form 
\bea
\x= \x^A D_A = \x^a \pa_a +\x^\a_{ij} D^{ij}_\a~, \qquad \x^\a_{ij} = \x^\a_{ji}~,
\eea
where $i, j $ are $\sSU(2)$ $R$-symmetry indices.
Requiring $\x$ to be conformal Killing, eq.  \eqref{def},  
and making use of the anti-commutation relation 
$\{D_\a^{ij},D_\b^{kl}\}=-2\ri \ve^{i(k}\ve^{l)j}\partial_{\a\b}$,  we 
deduce that
\begin{subequations} \label{3.25}
\bea
&&\x^{ij}_\a = -\frac{\ri}{6} D^{\b ij} \x_{\a\b}~, 
\label{2.10} \\
&& D_{(\a}^{ij}\x_{\b\g)}=0~.
\label{3.25b}
\eea
\end{subequations}
Here \eqref{3.25b} is the $\cN=3$ superconformal Killing equation. An important 
consequence one may derive from \eqref{3.25} is the identity 
\bea
D_\a^{(ij} \x_\b^{kl)} &=&0~.
\eea

The $\cN=3$ conformal supercurrent $J^\a$ satisfies
\bea
D_\a^{ij} J^{\a}=0~.
\label{3DN3supercurrent}
\eea
Let us define a real $\sSU(2)$ triplet $L^{ij} = L^{ji}$ associated with $J^\a$ and $\x^A$ 
by the rule:
\bea
L^{ij} = \ri \,\x_\a^{ij} J^\a +\frac{1}{4} \x^{\a\b} D^{ij}_\a J_\b~.
\eea
The properties of $J^\a$ and $\x^A$ imply that $L^{ij}$ is a linear 
multiplet, 
\bea
D_\a^{(ij}L^{kl)} =0~,
\eea
and therefore $L^{ij}$ contains a conserved current. 

Here we do not discuss the component content of $J^\a$ and $L^{ij}$. 
It can be readily determined, e.g., by making use of the $\cN=3 \to \cN=2$ superfield 
reduction of the $\cN=3 $ supercurrent and conserved current multiplets
described in \cite{BKS1}. We only point out that the conserved current, 
which is contained in $L^{ij}$, is given by 
\bea
V_{\a\b} = \ri \ve_{kl} D^{ik}_\a D_\b^{jl} L_{ij}| =V_{\b\a}~.
\eea


\subsection{$\cN=4$ superconformal symmetry}

Given an $\cN=4$ conformal Killing supervector field
\bea
\x = \x^A D_A = \x^a \pa_a + \x^\a_{i \bar i} D_\a^{i \bar i}~, 
\eea
it follow from \eqref{def} that 
\begin{subequations}
\bea
&&\x^{\a i \bar i} = \frac{\ri }{6} D_\b^{i \bar i} \x^{\b \a}~, \\
&&D^{i \bar i}_{(\a} \x_{\b \g)}=0~.
\label{3.32b}
\eea
\end{subequations}
Here $D_\a^{i \bar i}$ is the $\cN=4 $ spinor covariant derivative defined as in \cite{KPT-MvU}, 
with the  two-component indices $i $ and $\bar i $ corresponding to the 
left and right subgroups
of the $R$-symmetry group $\sSU(2)_{\rm L} \times \sSU(2)_{\rm R}$, respectively.
Eq. \eqref{3.32b} is the $\cN=3$ superconformal Killing equation.

The $\cN=4$ conformal supercurrent $J$ satisfies the conservation equation 
\bea
\ve^{\a\b} D_\a^{(i (\bar i} D_\b^{j) \bar j ) }J=0~.
\label{3DN4supercurrent}
\eea
Associated with $\x^A$ and $J$ is a left $\sSU(2)$ triplet 
$L^{ij} =L^{ji}$ defined by 
\bea
L^{ij} = \frac{\ri}{4}  \x^{\a\b} D_\a^{i \bar k} D_\b^j{}_{\bar k} J
+ \x^{\a (i \bar k} D_\a^{j)}{}_{\bar k} J
+ \L^{ij} J ~,
\eea
where we have introduced \cite{KPT-MvU}
\bea
\L^{ij} = \frac{1}{4} D^{\a (i}{}_{\bar k}  \x_\a^{j) \bar k} ~, 
\qquad 
D_\a^{ (i \bar i} \L^{jk)}=0~.
\eea
The properties of  $\x^A$ and $J$ imply that $L^{ij}$ is a left linear multiplet, 
\bea
D_\a^{(i \bar i} L^{jk)}=0~,
\eea
and therefore $L^{ij}$ contains a conserved current. 

In complete analogy with $L^{ij}$, one can also introduce
a right $\sSU(2)$ triplet 
$L^{\bar i \bar j} =L^{\bar j \bar i}$; it also contains a conserved current.

Here we do not discuss the component content of $J$ and $L^{ij}$. 
It can be readily determined, e.g., by making use of the $\cN=4\to \cN=3$ superfield 
reduction of the $\cN=4 $ supercurrent and conserved current multiplets
described in \cite{BKS1,BKS2}. We only point out that the conserved current, 
which is contained in $L^{ij}$, is given by 
\bea
V_{\a\b} = \ri \ve_{\bar i \bar j} D^{i \bar i }_\a D_\b^{j \bar j} L_{ij}| =V_{\b\a}~.
\eea


\section{Superconformal theories in four dimensions} 

In four dimensions, we consider only the $\cN=1$ and $\cN=2$
superconformal theories, because these cases
allow the existence of  conserved current multiplets without higher spin 
fields \cite{HST81}.

\subsection{$\cN=1$ superconformal symmetry}

Consider an arbitrary $\cN=1$ conformal Killing supervector field,
\bea
\x  = \x^a  \pa_a + \x^\a  D_\a
+ {\bar \x}_\ad  {\bar D}^\ad~.
\eea   
Its components are constrained according to \eqref{2.6ab}.
Let $J_{\a\ad}$ be the $\cN=1$ conformal supercurrent. 
It is a primary real vector superfield of dimension $+3$, as discussed, e.g., 
in \cite{Osborn}.  The supercurrent conservation equation  is 
given by eq. \eqref{1.5}.
Then the real scalar 
\bea
L= -\hf\x^{\ad \a} J_{\a \ad} 
\label{4.55}
\eea
 is a conserved current multiplet, 
\bea
D^2 L =0 \quad  \Longleftrightarrow \quad \bar D^2 L=0~.
\label{1.777}
\eea

It follows from  \eqref{1.5} that the conformal supercurrent has 
four independent components, which can be chosen as follows:
\bea
j_{\a\ad}:= J_{\a\ad}|~, \qquad S_{\a \b \ad} := D_\b J_{\a\ad}| =S_{(\a\b)\ad}~, 
\qquad T_{\a\b \ad \bd } := [D_{(\b}, \bar D_{( \bd}] J_{\a) \ad)} |~,
\eea
as well as the complex conjugate of $S_{\a \b \ad}$, $\bar S_{\a \ad \bd}$. 
Here $j_{\a\ad}$ is the $R$-symmetry current, $S_{\a \b \ad}$ and $\bar S_{\a \ad \bd}$
 the supersymmetry currents, and $T_{\a\b \ad \bd } = T_{(\a\b)  ( \ad \bd )}$
the energy-momentum tensor.\footnote{The definition of the energy-momentum tensor 
given in section 5.7.3 of \cite{BK} contains an error.} 
All these currents are conserved, 
\bea
\pa^{\ad \a} j_{\a\ad} = 0 ~, \qquad \pa^{\ad \a} S_{\a \b\ad} = 0 ~, \qquad
\pa^{\ad \a} T_{\a \b \ad \bd} = 0 ~,
\eea
as a consequence of \eqref{1.5}. 
We point out that the energy-momentum $T^{ab}$ is automatically traceless and the four-component Majorana supersymmetry current is $\g$-traceless.

Here we do not list all the component fields of $L$. 
We only point out that the conserved current contained in $L$ is given by 
\bea
V_{\a\ad} := [D_\a, \bar D_\ad ] L|~.
\eea
In order to compute the conserved current contained in \eqref{4.55}, 
it is necessary to make use of the explicit expression for the most general 
$\cN=1$ conformal Killing supervector field, which is given, e.g., in \cite{BK,Osborn}.


\subsection{$\cN=2$ superconformal symmetry}

Consider an arbitrary $\cN=2$ conformal Killing supervector field,
\bea
\x  = \x^a  \pa_a + \x^\a_i  D^i_\a
+ {\bar \x}_\ad^i  {\bar D}^\ad_i~.
\eea   
Its components are constrained according to \eqref{2.6ab}.
Let $J$ be the $\cN=2$ conformal supercurrent. As discussed in \cite{KT},
$J$ is a  primary real scalar superfield of dimension $+2$ and obeys
the conservation equation \eqref{sccl}.
We introduce the following real $\sSU(2)$ triplet
\bea
L^{ij}= \frac{\ri}{8} \x^{\ad \a} [D^{(i}_\a ,{\bar D}_{\ad}^{ j)}] J  
-   \L^{ij} J +  \big(
\x^{\a (i} D^{j)}_\a +\bar \x^{ \ad (i} \bar D^{j)}_\ad \big)J~.
\label{4.66}
\eea
Here  the real $\sSU(2)$ triplet  ${\L}^{i j}$ is defined as 
\bea
{\L}^{ij}  = -\frac{\ri}{32}
[D^{(i}_\a,{\bar D}_{\ad }^{j)}] \x^{\a\ad}
\eea
and has the properties \cite{KT}
\bea
D^{(i}_\a {\L}^{jk)} = 0 \quad \Longleftrightarrow \quad  {\bar D}^{(i}_\ad {\L}^{jk)}= 0~.
\eea
One can check that $L^{ij}$ is a linear multiplet, 
\be
D^{(i}_\a {L}^{jk)} = {\bar D}^{(i}_\ad {L}^{jk)}= 0~,
\ee
and therefore it contains a conserved current. 

The component content of $J$ and $L^{ij}$ is discussed, e.g.,  in \cite{KT}. 


\section{Superconformal  theories in five and six dimensions} 

The unique feature of five spacetime dimensions is that there is only one superconformal algebra \cite{Nahm}, which is isomorphic to the exceptional superalgebra $\rm F^2(4)$  and corresponds to the supersymmetry type $\cN=1$ with eight supercharges. Our 5D superspace notation and conventions correspond to \cite{KL} with one exception. 
Instead of using  Greek letters with a hat, such as $\hat \a, \hat \b$, 
for the four-component  spinor indices  \cite{KL}, here such indices will 
be denoted by ordinary Greek letters. 

Any real supervector field $\xi$ on 5D $\cN=1$ 
Minkowski superspace ${\mathbb M}^{5|8}$
has the form 
\bea
\x = \x^A D_A = \x^{ a}  \pa_{a} + \x^{\a}_i  D_{ \a}^i ~, \qquad i=1,2~, 
\eea
where $\x^a$ is real and $\x^\a_i$ obeys the pseudo-Majorana condition defined in
Appendix A of \cite{KL}. 
Requiring $\x$ to be conformal Killing, eq. \eqref{def},  
one obtains  \cite{K-compactified} (see also \cite{KNT-M14}) 
\begin{subequations} \label{5.2}
\bea
\x^i_{\a}&=&
\frac{\ri}{10}D^{\b i} \x_{\b \a}
~, 
\label{5.2a}
\\
D^i_{( \a} \x_{ \b)  \g} 
&=&-\frac{1}{5}  \,D^{  \d  i} \, \x_{ \d (  \a}  \, \ve_{ \b )  \g}~,
\label{5.2b}
\eea
\end{subequations}
where the traceless antisymmetric  rank-two spinor
$\x_{\a\b}$ is obtained from $\x^a$ by the standard rule
$\x_{\a\b} = (\G_c)_{\a\b} \x^c $, with the $\G$-matrices defined as in \cite{KL}. 
Eq. \eqref{5.2b} is the 5D superconformal Killing equation. 
One can deduce from \eqref{5.2} the following identities:
\bea
D_{ \a}^{(i} \, \x_{ \b}^{ j) } &=&\frac{1}{ 4} \,
\ve_{ \a  \b} D^{ \g (i }  \x_{ \g}^{ j)} 
\quad \Longrightarrow \quad 
D_{\a}^{(i} D_{\b}^{j} \x_{\g}^{k)} =0~, \qquad
(\G^{ b})_{ \a  \b} \,D^{ \a i} \x^{ \b }_i
=0~.
\label{master6}
\eea

The $\cN=1$ and $\cN=2$ supercurrents in five dimensions were introduced 
by Howe and Lindstr\"om \cite{HL81}. The conformal supercurrent, $J$,
is a  primary real scalar superfield of dimension $+3$, which obeys the conservation 
equation
 \cite{KNT-M14} 
 \bea
 D^{\a (i}D_\a^{j)} J =0 \quad \Longrightarrow \quad 
 D_\a^{(i} D_\b^j D_\g^{k)} J =0~.
 \label{5Dconservation}
 \eea
 Given a conformal Killing supervector field $\x^A$, we consider the following descendant of 
 the supercurrent: 
 \bea
 L^{ij}=\frac{\ri}{8}\x^{\a\b}D_\a^{(i} D_\b^{j)} J-\x^{\a (i} D_\a^{j)} J+\L^{ij} J ~,
\label{5.55}
 \eea
 where $\L^{ij}$ is defined by 
 \bea
 {\L}^{ij} = \frac{1}{ 4} D_{ \a }^{( i} \x^{j)  \a}
 \eea
and obeys the constraint 
\bea
D^{(i}_\a \L^{jk)}=0~.
\eea
Making use of the identities \eqref{5.2} and \eqref{master6} 
and the conservation equation \eqref{5Dconservation}, one may check that 
$L^{ij}$ is a linear multiplet, 
\bea
D^{(i}_\a L^{jk)}=0~,
\label{5.88}
\eea
and therefore it contains a conserved current. 

The expressions \eqref{4.66} and \eqref{5.55} look very similar. 
This feature is not accidental and actually it follows from the fact that 
the 4D $\cN=2$ and 5D $\cN=1$ supersymmetries describe eight supercharges. 
Another case with eight supercharges is the 6D $\cN=(1,0)$ supersymmetry, 
to which the above 5D analysis extends almost without changes. 
The only difference between the 5D and  6D cases is that 
the 6D $\cN=(1,0)$ conformal supercurrent, $J$,
is a  primary real scalar superfield of dimension $+4$, which obeys the conservation equation 
 \cite{HSierraT} 
\bea
D_\a^{(i} D_\b^j D_\g^{k)} J =0~,
\label{5.99}
 \eea
 which differs from the 5D conservation equation \eqref{5Dconservation}.
 However, the only property of the 5D supercurrent, which was crucial  
 in order to establish \eqref{5.88}, was the relation on the right hand side of  \eqref{5Dconservation}.
 The latter is the 5D counterpart of the 6D conservation equation \eqref{5.99}.


\section{Non-conformal supercurrents, curved backgrounds}

In this paper, we have presented the supersymmetric extensions 
of the construction $(T^{ab}, \x^a) \to V^a$, where 
$T^{ab}$ is the conserved and traceless energy-momentum tensor, 
$\x^a$ is an arbitrary conformal Killing vector field, and 
$V^a$ is the conserved current defined by \eqref{1.300}. 
As is well known, the field-theoretic 
construction has a simple modification to the non-conformal case
when $T^{ab}$ is no longer traceless, 
\bea
T^{ab}= T^{ba}~, \qquad  \pa_b T^{ab}=0~.
\eea
The vector field $V^a$ defined by \eqref{1.300} is still conserved provided $\x^a$ 
is a Killing vector field, 
\bea
\pa_a \x_b + \pa_b \x_a =0~.
\eea
This non-conformal construction also admits a supersymmetric generalisation. 
We will describe it only in the 4D case. To start with, we will briefly recall the structure 
of $\cN=1$ and $\cN=2$ non-conformal supercurrents.

A general non-conformal $\cN=1$ supercurrent 
is naturally associated with the non-minimal off-shell formulation 
\cite{Breitenlohner,SG} for $\cN=1$ supergravity.
The supercurrent conservation equation  (see, e.g., \cite{GGRS}) is 
\bea
{\bar D}^{\ad}{J} _{\a \ad} = a {\bar D}^2 
\z_\a
-b D_\a 
{\bar D}_\bd {\bar \z}^\bd 
~, \qquad D_{(\a}\z_{\b )} =0~,
\label{conservation-non}
\eea
with $a,b$ real parameters. Setting $\z_\a = D_\a Z$ leads to the supercurrent multiplet 
derived in \cite{MSW} using a version of 
the superfield Noether procedure elaborated in \cite{Osborn}.

An alternative form for the general $\cN=1$ supercurrent, 
which is  simply related to \eqref{conservation-non},
was presented in \cite{K-var}. It naturally 
follows from the classification of the linearised $\cN=1$ supergravity actions 
given in \cite{GKP} and is described by the  conservation equation 
\begin{subequations} \label{6.2}
\bea
& {\bar D}^{\ad}{J} _{\a \ad} = { \c}_\a  +{\rm i}\,\eta_\a +D_\a X~, &
 \\ 
& {\bar D}_\ad {\c}_\a  = {\bar D}_\ad \eta_\a= {\bar D}_\ad {X}=0~, 
\qquad D^\a {\c}_\a - {\bar D}_\ad {\bar {\c}}^\ad
=D^\a {\eta}_\a - {\bar D}_\ad {\bar {\eta}}^\ad = 0~. &
\label{VIII}
\eea
\end{subequations}
The chiral superfields 
$\c_\a$, $\eta_\a$ and $X$ constitute the so-called multiplet of anomalies. 
In principle, one may always solve the constraints imposed 
on $\c_\a$, $\eta_\a$ and $X$
in terms of unconstrained  potentials as follows 
\bea
\c_\a = -\frac{1}{4} {\bar D}^2 D_\a V~, \qquad 
\eta_\a =  -\frac{1}{4} {\bar D}^2 D_\a U~, \qquad 
X = -\frac{1}{4} {\bar D}^2 Z~,
\eea
where $V$ and $U$ are real. However, in some cases this is accompanied by the loss
of locality (that is, some of the potentials are not well-defined local operators)
and gauge invariance. This point of view was advocated in \cite{KS2}.
The supercurrent \eqref{6.2} with 
$\c_\a = \eta_\a =0$ was introduced by Ferrara and Zumino \cite{FZ}, and it is associated with 
the old minimal formulation \cite{old-WZ,old} for $\cN=1$ supergravity.
The supercurrent \eqref{6.2} with $ \eta_\a =0 $ and $X=0$  corresponds to 
the new minimal formulation \cite{new} for $\cN=1$ supergravity.
Sometimes it is called the $R$-multiplet \cite{KS2}. 
The supercurrent \eqref{6.2} with $ \c_\a =0 $ and $X=0$  corresponds to 
the exotic minimal supergravity formulation \cite{BGLP}, which is known only at the 
linearised level. This supercurrent is sometimes called the virial multiplet 
\cite{Nakayama}. 
The supercurrent \eqref{6.2} with $ \eta_\a =0 $ is known as the $S$-multiplet
\cite{KS2}. It does not correspond to any irreducible supergravity theory, 
although it was argued \cite{KS2} to be universal in the case of 
$\cN=1$ Poincar\'e supersymmetry.\footnote{The $S$-multiplet does not exist 
in the case of $\cN=1$ anti-de Sitter supersymmetry \cite{ButterK2}, 
for which the Ferrara-Zumino supercurrent is universal.} 

Let us also reproduce a non-conformal deformation of the 
$\cN=2$ supercurrent multiplet \eqref{sccl} that supports a large family of 
$\cN=2$ supersymmetric field theories. The corresponding conservation equation 
\cite{ButterK2,ButterK1} is
\begin{subequations}
\begin{align}\label{eq_N2current}
\frac{1}{4} D^{ij} J = w T^{ij} - g^{ij} Y~,
\end{align}
where $T^{ij}$ and $Y$ are the trace multiplets constrained by 
\begin{gather}\label{eq_CurrentConstraints}
D_\alpha^{(i} T^{jk)} = \bar D_\ad^{(i} T^{jk)} = 0 ~,\qquad \overline{T^{ij}} =T_{ij}~,
\\
\bar D_\ad^i Y =0~, \qquad D^{ij} Y = \bar D^{ij} \bar Y~,
\end{gather}
\end{subequations}
The right-hand side of  \eqref{eq_N2current}
involves two constant parameters, complex $w$ and real $\sSU(2)$ triplet 
$g^{ij}$, which may be thought of as expectation values of the 
two conformal compensators in the off-shell formulations for $\cN=2$ supergravity 
developed by de Wit, Philippe and Van Proeyen \cite{deWPV}.
The supercurrent multiplet with $g^{ij} =0$ is equivalent to the one 
discovered originally by Sohnius \cite{Sohnius}.

In the remainder of this section, our analysis will be restricted 
to the  $\cN=1$ case and only the Ferrara-Zumino supercurrent  \cite{FZ} 
will be studied (all technical steps are analogous for the other supercurrents).
The corresponding conservation equation is 
\bea
\bar D^\ad J_{\a\ad}= D_\a X~, \qquad \bar D_\ad X=0~,
\label{6.55}
\eea
with $X$ the chiral trace multiplet.\footnote{Since 
$D^2 X - \bar D^2 \bar X = -2\ri \pa_{\a\ad} J^{\ad \a}$, 
the chiral trace $X$ in \eqref{6.55} is in fact an example of the three-form multiplet \cite{Gates,GGRS}.} 
If $X\neq 0$, the real scalar $L$ defined by \eqref{4.55} is no longer a 
linear superfield. 
Conservation equation  \eqref{1.777}
turns into
\bea
\bar D^2 L = 2\ri \x X = 2\ri (\x^a \pa_a +\x^\a D_\a)X~.
\label{6.8}
\eea
Here the right-hand side is chiral, because $\x X $ is the variation of 
the chiral superfield $X$ generated by 
the conformal Killing supervector field  $\x =\x^A D_A$.
If $\x$ is a Killing supervector field, then it obeys 
the additional constraint \cite{BK} 
\bea
D_\a \x^\a = \bar D_\ad \bar \x^\ad =0 \quad \Longrightarrow \quad 
\pa_a \x^a =0~.
\eea
In the case that $\x$ is Killing, the  relation \eqref{6.8}
is equivalent to 
\bea
\bar D^2 L = 2\ri \x X = 2\ri \Big\{ \pa_a (\x^a X)  -D_\a (\x^\a X)\Big\}~.
\label{6.3}
\eea
Since $\bar D^{(\ad} \x^{\bd) \b} =0 $, eq. \eqref{2.6b}, we can represent 
\bea
\x_{\a\ad} = -2\ri \bar D_\ad \U_\a \quad 
\Longrightarrow \quad \x_\a = -\frac{1}{4} \bar D^2 \U_\a~,
\eea
for some spinor $\U_\a$. Making use of this representation, eq. \eqref{6.3}
may be rewritten in the form 
\bea
\bar D^2 \tilde L = 0~, 
\qquad \tilde L : = L +\frac{\ri}{2} \Big\{ D^\a(\U_\a X) - \bar D_\ad (\bar \U^\ad \bar X) 
\Big\}~.
\eea
We conclude that $\tilde L$ contains a conserved current. 

So far, our discussion in this paper has been restricted to theories
in flat superspace. However, practically all considerations and conclusions 
may be extended
to supersymmetric field theories defined on curved superspace backgrounds
with symmetries. 
As an example, let us consider a curved superspace background $\cM^{4|4}$ of the 4D $\cN=1$ old minimal supergravity.\footnote{Our supergravity conventions follow \cite{BK} 
and slightly differ from those used in \cite{WB}.}  
Let $J_{\a\ad}$ be the conformal supercurrent, 
\bea
\cD^\a J_{\a\ad} =0 \quad \Longrightarrow \quad (\cD^2 - 6 \bar R ) J_{\a\ad}=0~.
\eea
Let $\x= \x^A E_A $ be a conformal Killing supervector field of $\cM^{4|4}$. 
As demonstrated in section 6.4 of  
 \cite{BK},  its explicit form is 
 \bea
\x^A = \Big( \x^a, \x^\a, \bar \x_\ad \Big) 
= \Big( \x^a, -\frac{\ri }{8} \bar \cD_\bd \x^{\bd \a} , 
-\frac{\ri}{8} \cD^\b \x_{\b \ad} \Big)
\eea
where the vector component $\x_{\a\ad} $ is real and obeys the equation
\bea
\cD_{(\a} \x_{\b) \bd} =0  \quad \Longrightarrow \quad (\cD^2 +2 \bar R ) \x_{\a\ad}=0~.
\label{6.155}
\eea
Then the real scalar $L:= -\hf\x^{\a\ad} J_{\a \ad} $ is a conserved current multiplet,
\bea
(\cD^2 -4\bar R) L =0~.
\eea


\noindent
{\bf Acknowledgements:}\\
The work of SMK and ST  is supported
in part by the Australian Research Council (ARC) Discovery Project DP140103925.

\appendix 

\section{Conformal symmetries of curved superspace}\label{AppendixA}

The material in this section is taken almost verbatim from \cite{K15Corfu}.

Let $\cM^{d|\d}$ be a curved superspace, with $d$ spacetime 
and $\d$ fermionic dimensions, chosen to describe a given supergravity theory.
We denote by  $z^M = (x^m, \q^{\hat \m}) $ the local coordinates
for  $\cM^{d|\d}$. Without loss of generality, we assume that 
the zero  section of $\cM^{d|\d}$  defined by $\q^{\hat \m} =0$ corresponds to 
the spacetime manifold $\cM^d$. 

The  differential geometry of curved superspace $\cM^{d|\d}$ may be realised 
in terms of covariant derivatives  of the form 
\bea
\cD_A = (\cD_a, \cD_{\hat \a}) = E_A +\O_A +\F_A ~.
\label{3.1}
\eea
Here $E_A = E_A{}^M (z)\pa / \pa z^M $ denotes the inverse superspace vielbein,
$\O_A = \hf \O_A{}^{bc} (z) M_{bc} $ is the  Lorentz connection, 
and $\F = \F_A{}^I(z) J_I $ the $R$-symmetry connection.\footnote{The superspace 
structure group,  ${\rm Spin}(d-1,1) \times G_R$, is 
a subgroup of the isometry group of Minkowski superspace ${\mathbb M}^{d|\d}$. 
This subgroup is the isotropy group of the origin in ${\mathbb M}^{d|\d}$.}
The index $\hat \a$ of the fermionic operator $\cD_{\hat \a}$ is, in general,  composite; 
it is comprised of a spinor index $\a$  and an $R$-symmetry index. 

The covariant derivatives obey the (anti-)commutation relations of the form
\bea
[ \cD_A , \cD_B\} = \cT_{AB}{}^C \cD_C + \hf \cR_{AB}{}^{cd} M_{cd} 
+\cR_{AB}{}^I J_I~, 
\label{3.2}
\eea
where $ \cT_{AB}{}^C (z)$ is the torsion tensor, $ \cR_{AB}{}^{cd} (z)$ and 
$\cR_{AB}{}^I (z)$ are the Lorentz and $R$-symmetry curvature tensors, respectively. 
In order to describe conformal supergravity, the superspace torsion $\cT_{AB}{}^C$
has to obey certain algebraic constraints.

The supergravity gauge group includes a subgroup  generated by local transformations
\bea
\d_\cK \cD_A &=& [\cK , \cD_A] ~, \qquad 
\qquad \cK:= \x^B (z) \cD_B + \hf K^{bc} (z) M_{bc} + K^I (z) J_I ~, 
\label{3.3a}
\eea
where the gauge parameters $\x^B$, $K^{bc}= -K^{cb}$ and $K^I$
obey standard reality conditions but are  otherwise  arbitrary. 

In order to describe conformal supergravity, the constraints imposed on the superspace 
torsion should be invariant under super-Weyl transformations \cite{HT} of the form
\bea
\d_\s \cD_a &=& \s \cD_a + \cdots~, \qquad \d_\s \cD_\hal = \hf \s \cD_\hal + \cdots~, 
\label{3.4a}
\eea
where the scale parameter $\s$ is an arbitrary real superfield. The ellipsis 
in the expression for $\d_\s \cD_a $ includes, in general, 
a linear combination of the spinor covariant derivatives $\cD_{\hat \b}$ and 
the structure group generators $M_{cd}$ and $J_K$.  The ellipsis in 
$\d_\s \cD_\hal $ stands for a linear combination of the generators
of  the structure group. 
Consider the superspace vielbein $E^A = \rd z^M E_M{}^A(z) $ to which $E_A$ is dual. 
The specific feature of the super-Weyl transformation is that the vector one-form $E^a$ 
transforms homogeneously, 
\bea
\d_\s E^a = - \s E^a~.
\eea
This implies that every super-Weyl 
transformation at most scales 
the superspace interval defined by $\rd s^2 := \eta_{a b} E^a E^b$.
The  Lorentz and $R$-symmetry transformations preserve the interval.

Let us now fix  a background superspace. 
A supervector field $\x= \x^B E_B$  on $(\cM^{d |\d}, \cD)$ is called 
conformal Killing if 
\bea
 (\d_\cK + \d_\s) \cD_A = 0~,
 \label{3.5}
\eea
for some Lorentz $K^{bc}$, $R$-symmetry $K^I$ and super-Weyl 
$\s$ parameters. For any dimension $d\leq 6$ and any conformal 
supergravity with up to eight supercharges, the following properties hold:
(i) all parameters $K^{bc} $,  $K^I$ and $\s$ are uniquely determined in terms of $\x^B$, which allows us to write $K^{bc} = K^{bc}[\x]$, $K^I = K^I[\x]$ and 
$\s = \s[\x]$;
(ii) 
the spinor component $\x^{\hat \b}$ is uniquely determined in terms of $\x^b$;
(iii) the vector component $\x^b$ obeys 
a superconformal Killing equation, which contains all the information 
about the  conformal Killing vector field and, in particular, 
implies the ordinary conformal Killing equation 
\bea
\cD_a \x_b + \cD_b \x_a = \frac{2}{d} \eta_{ab} \cD_c \x^c~.
\label{A.7}
\eea 
Unlike \eqref{A.7}, the explicit form of the superconformal Killing equation 
depends on the spacetime dimension
and supersymmetry type chose. 
For instance, in the case of 4D $\cN=1$ supergravity 
this equation  \cite{BK} is given by \eqref{6.155}.


\begin{footnotesize}

\end{footnotesize}

\end{document}